\documentclass[twocolumn]{revtex4}

\usepackage{graphicx}% Include figure files
\usepackage{dcolumn}% Align table columns on decimal point
\usepackage{bm}% bold math

\input epsf

\begin{document}

\title{\Large Thermodynamical Laws in Ho$\check{\text r}$ava-Lifshitz Gravity}

\author{\bf Samarpita Bhattacharya\footnote{samarpita$_{-}$sarbajna@yahoo.co.in} and Ujjal Debnath\footnote{ujjaldebnath@yahoo.com ,
ujjal@iucaa.ernet.in}}

\affiliation{Department of Mathematics, Bengal Engineering and
Science University, Shibpur, Howrah-711 103, India.}

\date{\today}

\begin{abstract}
In this work, we have investigated the validity of GSL of
thermodynamics in a universe (open, closed and flat) governed by
Ho$\check{\text r}$ava-Lifshitz gravity. If the universe contains
barotropic fluid the corresponding solutions have been obtained.
The validity of GSL have been examined by two approaches: (i)
robust approach and (ii) effective approach. In robust approach,
we have considered the universe contains only matter fluid and the
effect of the gravitational sector of HL gravity was incorporated
through the modified black hole entropy on the horizon. Effective
approach is that all extra information of HL gravity into an
effective dark energy fluid and so we consider the universe
contains matter fluid plus this effective fluid. This approach is
essentially same as the Einstein's gravity theory. The general
prescription for validity of GSL have been discussed. Graphically
we have shown that the GSL may be satisfied for open, closed and
flat universe on the different horizons with different
conditions.\\
\end{abstract}


\maketitle

\section{\normalsize\bf{Introduction}}

Recently, a power-counting renormalizable theory of gravity was
proposed by Ho$\check{\text r}$ava [1-4]. This is a
non-relativistic theory of gravity and is expected to recover
Einstein's general relativity at large scales. This theory does
not have the full diffeomorphism invariance. Although presenting
an infrared (IR) fixed point, namely General Relativity, in the UV
the theory exhibits an Lifshitz type anisotropic scaling between
time and scale, so it is commonly known as Ho$\check{\text
r}$ava-Lifshitz (HL) gravity. The HL gravity theory has attracted
much attention as a candidate quantum field theory of gravity with
$z=3$ in the UV, where $z$ measures the degree of anisotropy
between space and time. In 3+1 dimensions, the Ho$\check{\text
r}$ava-Lifshitz theory has a $z=3$ fixed point in the UV and flows
to a $z=1$ fixed point in the IR, which is just the classical
Einstein-Hilbert gravity theory. In this theory the effective
coupling constant is dimensionless. This theory of gravity has
four possible versions so far - with/without the detailed balance
condition and with/without projectability condition. Among these
version without detailed balance and with the projectability
condition is the most viable one.\\

Ho$\check{\text r}$ava-Lifshitz gravity has been studied and
extended in detail [5] and it has been applied as the cosmological
framework of the universe [6, 7]. It was extensively applied to a
resolution of the cosmological problem including inflation and
non-Gaussianity in [8, 9], and new solutions were constructed in
[10, 11]. Amongst the very interesting physical implications are
the novel solution subclass [10,12], the gravitational wave
production [13], dark energy phenomenology [14], astrophysical
phenomenology [15], observational constraints [16] etc. Static
spherically symmetric black hole solutions in Ho$\check{\text
r}$ava- Lifshitz gravity has been studied in [10]. However,
despite this extended research, there are still many ambiguities
if Ho$\check{\text r}$ava-Lifshitz gravity is reliable and capable
of a successful description of the gravitational background of our
world, as well as of the cosmological behavior of the universe
[17]. There are several work on the thermodynamical properties of
black hole in Ho$\check{\text r}$ava-Lifshitz gravity theory [18].\\

Here we consider the universe is treated as a thermodynamical
system and we discuss the GSL of thermodynamics in Ho$\check{\text
r}$ava-Lifshitz gravity theory. Till now there are some works [19]
have been done on thermodynamics in FRW model with HL gravity.
Here we try to examine the validity of GSL on the horizons in
Ho$\check{\text r}$ava-Lifshitz gravity theory. For this purpose,
we investigate two approaches. The first approach is to consider
the FRW universe is filled with only matter field and using first
law, we examine the validity of GSL on the horizons and give some
conditions in HL gravity. Second approach is to absorb all the
extra terms of HL gravity in an dark energy component and we
consider the universe is filled with dark matter and this type of
dark energy. But the first approach is more interesting, because
we have directly get the additional information of HL gravity
model. Recently Jamil et al [20] have discussed the validity of
GSL considering these two types of approaches. We have extend
their work in different horizons like, Hubble, apparent, particle
and event horizons in general manner. Finally, we give some
solutions for barotropic fluid and justify
the validity of GSL in different horizons.\\

\section{\normalsize\bf{Ho$\check{\text r}$ava-Lifshitz Gravity Theory}}

We briefly review here the scenario where the cosmological
evolution is governed by Ho$\check{\text r}$ava-Lifsfitz gravity
[6, 7]. The dynamical variables are the lapse rate and shift
functions, $N$ and $N_{i}$ respectively, and the spatial metric
$g_{ij}$. In the (3+1) dimensional Arnowitt-Deser-Misner formalism
the full metric [21] is written as

\begin{equation}
ds^{2}=-N^{2}dt^{2} + g_{ij}(dx^{i} + N^{i}dt)(dx^{j} + N^{j}dt)
\end{equation}

The scaling transformation of the coordinates reads: $t\rightarrow
l^{3}t$ and $x^{i}\rightarrow lx^{i}$. The gravitational action is
decomposed into a kinetic and a potential part as $S=\int dt
d^{3}x \sqrt{g}N({\cal L}_{K} + {\cal L}_{V})$. Under the detailed
balance condition the full action condition of Ho$\check{\text
r}$ava-Lifshitz gravity is given by

\begin{eqnarray*}
S=\int dt d^{3}x \sqrt{g}N\left[\frac{2}{\kappa^{2}}(K_{ij}K^{ij}
- \lambda K^{2}) + \frac{\kappa^{2}}{2 \omega^{4}}C_{ij}C^{ij}
\right.
\end{eqnarray*}
\begin{eqnarray*}
~~~~~~~~~~~~- \frac{\kappa^{2}\mu \epsilon^{ijk}}{2 \omega^{2}
\sqrt{g}} R_{il}\nabla_{j}R^{l}_{k}+
\frac{\kappa^{2}\mu^{2}}{8}R_{ij}R^{ij}
\end{eqnarray*}
\begin{equation}
\left.~~~~~~~~~~~~~  + \frac{\kappa^{2}\mu^{2}}{8(3 \lambda -
1)}\left(\frac{1 -4 \lambda}{4}R^{2} + \Lambda R - 3 \Lambda^{2}
\right) \right]
\end{equation}

where

\begin{equation}
K_{ij}=\frac{1}{2N}(\dot{g}_{ij} - \nabla_{i}N_{j}-
\nabla_{j}N_{i})
\end{equation}

is the extrinsic curvature and

\begin{equation}
C^{ij}=\frac{\epsilon^{ikl}}{\sqrt{g}}\nabla_{k}(R^{j}_{i} -
\frac{1}{4}R \delta^{j}_{l})
\end{equation}

is known as Cotton tensor and the covariant derivatives are
defined with respect to the spatial metric $g_{ij}$.
$\epsilon^{ijk}$ is the totally antisymmetric unit tensor,
$\lambda$ is a dimensionless coupling constant and the variable
$\kappa$ , $\omega$ and $\mu$ are constants with mass dimensions
$-1,~ 0,~ 1$ respectively. Also $\Lambda$ is a positive constant,
which as usual is related to the cosmological constant in the IR
limit. In order to incorporate the (dark plus baryonic) matter
component one adds a cosmological stress-energy tensor to the
gravitational field equations, by demanding to recover the usual
general relativity formulation in the low energy limit. Let us
suppose the
energy density and pressure are denoted by $\rho$ and $p$ respectively.\\

Now, in order to focus on cosmological frameworks, we impose the
so called projectability condition [17] and use a
Friedmann-Robertson-Walker (FRW) metric,

\begin{equation}
N=1,    g_{ij}=a^{2}(t)\gamma_{ij},      N^{i}=0
\end{equation}

with

\begin{equation}
\gamma_{ij} dx^{i} dx^{j}= \frac{dr^{2}}{1 - kr^{2}}+
r^{2}d\Omega^{2}_{2},
\end{equation}
where $k=0, -1, +1$ corresponding to flat, open and closed
respectively. By varying $N$  and $g_{ij}$, we obtain the
non-vanishing equations of motions:

\begin{eqnarray*}
H^{2}=\frac{\kappa^{2}}{6(3\lambda -1)} ~ \rho +
\frac{\kappa^{2}}{6(3\lambda -1)} \left[\frac{3 \kappa^{2} \mu^{2}
k^{2}}{8(3\lambda -1)a^{4}} + \frac{3 \kappa^{2} \mu^{2}
\Lambda^{2}}{8(3\lambda -1)}\right]
\end{eqnarray*}
\begin{equation}
~~~~~~~~~~~~~~~~~~~~~~~~~~~~~~~~~~~ - \frac{ \kappa^{4} \mu^{2}k
\Lambda }{8(3\lambda -1)^{2}a^{2}}
\end{equation}
and
\begin{eqnarray*}
\dot{H} + \frac{3}{2}H^{2}= - \frac{\kappa^{2}}{4(3\lambda -1)}~p
- \frac{\kappa^{2}}{4(3\lambda -1)} \left[\frac{ \kappa^{2}
\mu^{2} k^{2}}{8(3\lambda -1)a^{4}}\right.
\end{eqnarray*}
\begin{equation}
\left.~~~~~~~~~~~~~~~ - \frac{3 \kappa^{2} \mu^{2}
\Lambda^{2}}{8(3\lambda -1)}\right] - \frac{ \kappa^{4} \mu^{2}k
\Lambda }{16(3\lambda -1)^{2}a^{2}}
\end{equation}

where $H\equiv\frac{\dot{a}}{a}$ is the Hubble parameter. The term
proportional to $a^{-4}$ is the usual ``dark radiation", present
in Ho$\check{\text r}$ava-Lifshitz cosmology while the constant
term is just the explicit cosmological constant. For $k=0$, there
is no contribution from the higher order derivative terms in the
action. However for $k\ne 0$, there higher derivative terms are
significant for small volume i.e., for small $a$ and become
insignificant for large $a$, where it agrees with general
relativity. As a last step, requiring these expressions to
coincide the standard Friedmann equations, in units where $c=1$ we
set [6, 7],

\begin{equation}
G_{c}=\frac{\kappa^{2}}{16 \pi(3\lambda -1)}
\end{equation}

\begin{equation}
\frac{ \kappa^{4} \mu^{2} \Lambda }{8(3\lambda -1)^{2}}=1
\end{equation}

where $G_{c}$ is the ``cosmological" Newton's constant. We mention
that in theories with Lorentz invariance breaking (such is
Ho$\check{\text r}$ava-Lifshitz one) the ``gravitational" Newton's
constant $G$, that is the one that is present in the gravitational
action, does not coincide with $G_{c}$ , that is the one that is
present in Friedmann equations, unless Lorentz invariance is
restored [22], where

\begin{equation}
G=\frac{\kappa^{2}}{32 \pi}
\end{equation}

as it can be straightforwardly read from the action (2). In the IR
$(\lambda=1)$ where Lorentz invariance is restored, $G_{c}=G$.
Using the above identifications, we can re-write the Friedmann
equations $(7)$ and $(8)$ as,

\begin{equation}
H^{2} + \frac{k}{a^{2}}=\frac{8\pi G_{c}}{3}~\rho + \frac{k^{2}}{2
\Lambda a^{4}} + \frac{\Lambda}{2}
\end{equation}
and
\begin{equation}
\dot{H} + \frac{3}{2}H^{2} + \frac{k}{2a^{2}}= -4 \pi G_{c}p -
\frac{k^{2}}{4 \Lambda a^{4}} + \frac{3\Lambda}{4}
\end{equation}

If we consider the matter is conserved then the continuity
equation is given by

\begin{equation}
\dot{\rho}+3H(\rho+p)=0
\end{equation}

In the next section, we will describe the general condition for
validity of GSL of thermodynamics on the Hubble, apparent,
particle and event horizons due to presence of HL gravity in FRW universe.\\

\section{\normalsize\bf{Generalized Second Law of Thermodynamics in FRW universe in Ho$\check{\text r}$ava-Lifshitz gravity}}

We consider the FRW universe in Ho$\check{\text r}$ava-Lifshitz
gravity as a thermodynamical system with the horizon surface as a
boundary of the system. To study the generalized second law (GSL)
of thermodynamics through the universe we deduce the expression
for normal entropy using the Gibb's equation of thermodynamics
[27]

\begin{equation}
T_{X}dS_{IX}=pdV+d(E_{X})
\end{equation}

where, $S_{IX},~p,~V$ and $E_{X}$ are respectively entropy,
pressure, volume and internal energy within the
Hubble/apparent/particle/event horizon and $T_{X}$ is the
temperature on the Hubble horizon ($X=H$)/apparent horizon
($X=A$)/particle horizon ($X=P$)/event horizon ($X=E$). Here the
expression for internal energy can be written as $E_{X}=\rho V$.
Now the volume of the sphere is $V=\frac{4}{3}\pi R_{X}^{3}$,
where $R_{X}$ is the radius of the Hubble horizon
($R_{H}$)/apparent horizon ($R_{A}$)/particle horizon
($R_{P}$)/event horizon ($R_{E}$) defined by [27, 34]

\begin{equation}
R_{H}=\frac{1}{H}~,
\end{equation}

\begin{equation}
R_{A}=\frac{1}{\sqrt{H^{2} +
\frac{k}{a^{2}}}}=\frac{1}{\sqrt{H^{2}+k(1+z)^{2}}},
\end{equation}

\begin{equation}
R_{P}=a\int_{0}^{t}\frac{dt}{a}=a\int_{0}^{a}\frac{da}{a^{2}H}=\frac{1}{1+z}\int_{z}^{\infty}\frac{dz}{H}
\end{equation}

and
\begin{equation}
R_{E}=a\int_{t}^{\infty}\frac{dt}{a}=a\int_{a}^{\infty}\frac{da}{a^{2}H}=\frac{1}{1+z}\int_{-1}^{z}\frac{dz}{H}
\end{equation}

where $z$ is the redshift defined by $z=\frac{1}{a}-1$. Now
differentiating (16) - (19), with respect to time $t$, we find

\begin{equation}
\dot{R}_{H}= - \frac{\dot{H}}{H^{2}}
\end{equation}

\begin{equation}
\dot{R}_{A}=HR_{A}^{3}\left(k(1+z)^{2} - \dot{H} \right)
\end{equation}

\begin{equation}
\dot{R}_{P}=HR_{P}+1
\end{equation}

and

\begin{equation}
\dot{R}_{E}=HR_{E}-1
\end{equation}

The temperature on the Hubble/ apparent/ particle/ event horizon
is chosen as

\begin{equation}
T_{X}=\frac{1}{2 \pi R_{X}}
\end{equation}

Now from (15) we obtain, the rate of change of internal entropy
(using $(15)$ and $(24)$) as,

\begin{equation}
\dot{S}_{IX}=8\pi^{2}R_{X}^{3}(\rho+p)( \dot{R}_{X} - H R_{X})
\end{equation}

In case of black holes in Ho$\check{\text r}$ava-Lifshitz gravity
and under the detailed balance condition, the expression for the
entropy of the horizon [11, 23] is given by

\begin{equation}
S_{X}=\frac{4 \pi^{2} \kappa^{2} \mu^{2}}{4}[\Lambda R_{X}^{2} +
2k~ln(\sqrt{\Lambda}R_{X})]
\end{equation}

which implies (after differentiating)

\begin{equation}
\dot{S}_{X}=\frac{2 \pi}{G}R_{X}\dot{R}_{X} + \frac{2 \pi k }{G
\Lambda R_{X} }\dot{R}_{X}
\end{equation}

where in IR limit $(\lambda=1)$ we get $G_{c}=G$. \\

Therefore rate of change of total entropy in HL gravity is
obtained as (adding (25) and (27))

\begin{eqnarray*}
\dot{S}_{IX}+\dot{S}_{X}=8 \pi^{2} R_{X}^{3}( \dot{R}_{X} - H
R_{X})(\rho+p)
\end{eqnarray*}
\begin{equation}
~~~~~~~~~~~~~~~~~~~~~~~~~+ \frac{2 \pi}{G}\left(R_{X} +
\frac{k}{\Lambda R_{X}}\right)\dot{R}_{X}
\end{equation}

The GSL in HL gravity will satisfy if the following condition
holds:

\begin{eqnarray*}
\dot{S}_{IX}+\dot{S}_{X}\ge 0~~i.e.,~~8 \pi^{2} R_{X}^{3}(
\dot{R}_{X} - H R_{X})(\rho+p)
\end{eqnarray*}
\begin{equation}
~~~~~~~~~~~~~~ + \frac{2 \pi}{G}\left(R_{X} + \frac{k}{\Lambda
R_{X}}\right)\dot{R}_{X} \ge 0
\end{equation}

From the above restrictions, we can not draw any definite
conclusions for validity of GSL in HL gravity for flat, open or
closed FRW universe on different horizons. On the apparent
horizon, Jamil et al [20] have found that for flat and closed
universe, GSL is always satisfied but for open universe, the GSL
may be satisfied for some conditions upon $\Lambda$. For $k=0$ we
get back similar to the simple Einstein's gravity. So we are
interested to get results only for $k\ne 0$. In this case, the
results for HL gravity will be obtained. In the next section,
we'll consider the matter fluid is followed by barotropic equation
of state. For this type of fluid, we'll find some solutions and it
is easy to verify
the validity of GSL on different horizons.\\

\section{\normalsize\bf{Validity of GSL on Hubble, apparent, Particle and
Event Horizons in the presence of barotropic fluid}}

Let us consider the equation of state for the barotropic fluid is

\begin{equation}
p=w\rho
\end{equation}

where $w$ is a constant. Now solving equation (14) we get the
expression for energy density $\rho$ in terms of redshift $z$ as

\begin{equation}
\rho=\rho_{0}(1 + z)^{3(1 + w)}
\end{equation}

where $\rho_{0}$ is integration constant. Solving $(12)$ and
$(13)$ for $H$ and $ \dot{H}$ in terms of redshift $z$, obtained
as

\begin{equation}
H=\left[\frac{8 \pi G_{c}}{3}\rho_{0}(1 + z)^{ 3(1 + w)} +
\frac{k^{2}}{2 \Lambda }(1 + z)^{4} + \frac{\Lambda}{2} - k(1
+z)^{2}\right]^{\frac{1}{2}}
\end{equation}
and
\begin{equation}
\dot{H}= - 4 \pi G_{c}\rho_{0}(1 + w)(1 +z)^{3 (1+w) }  -
\frac{k^{2}}{ \Lambda }(1 + z)^{4}+ k(1 + z)^{2}
\end{equation}

Also the radii of particle and event horizons (from (18) and (19))
can be written as

\begin{eqnarray*}
R_{P}=\frac{1}{1+z}\int_{z}^{\infty}\left[ \frac{8 \pi
G_{c}}{3}\rho_{0}(1 + z)^{ 3(1 + w_{M})} + \frac{k^{2}}{2 \Lambda
}(1 + z)^{4} \right.
\end{eqnarray*}
\begin{equation}
~~~~~~~~~~~~~~~~~~~~~~~~~~~\left. + \frac{\Lambda}{2} - k(1
+z)^{2}\right]^{-\frac{1}{2}}dz
\end{equation}
and
\begin{eqnarray*}
R_{E}=\frac{1}{1+z}\int_{-1}^{z}\left[ \frac{8 \pi
G_{c}}{3}\rho_{0}(1 + z)^{ 3(1 + w_{M})} + \frac{k^{2}}{2 \Lambda
}(1 + z)^{4} \right.
\end{eqnarray*}
\begin{equation}
~~~~~~~~~~~~~~~~~~~~~~~~~~~\left. + \frac{\Lambda}{2} - k(1
+z)^{2}\right]^{-\frac{1}{2}}dz
\end{equation}

Let us now proceed to calculation of total entropy variation with
respect to time $t$.\\

Using equations (20) and (28), the total entropy variation on
Hubble horizon is obtained as

\begin{eqnarray*}
\dot{S}_{IH}+\dot{S}_{H}=\frac{2
\pi}{GH^{3}}\left(\frac{\dot{H}}{H^{2}} + 1\right)\left[\dot{H} -
k(1 + z)^{2}\right.
\end{eqnarray*}
\begin{equation}
\left. ~~~~~~~~~~~~ + \frac{k^{2}(1 + z)^{4}}{\Lambda }\right] -
\frac{2 \pi \dot{H}}{G H^{2}} \left(\frac{1}{H} + \frac{k
H}{\Lambda}\right)
\end{equation}

Using equations (21) and (28), the total entropy variation on
Apparent horizon is obtained as

\begin{eqnarray*}
\dot{S}_{IA}+\dot{S}_{A}=\left[\frac{2 \pi H
R_{A}^{6}}{G}\left(\dot{H} + k(1 + z)^{2} + \frac{k^{2}(1 +
z)^{4}}{\Lambda }\right) \right.
\end{eqnarray*}
\begin{equation}
\left.~~~~~ - \frac{2 \pi k}{ G \Lambda}R_{A}^{2}H\right]\times
\left(\dot{H} -  k(1 + z)^{2}\right)+ \frac{2 \pi k^{2}}{ G
\Lambda}R_{A}^{4}H(1 +z)^{4}
\end{equation}

\begin{figure}
\includegraphics[height=2in]{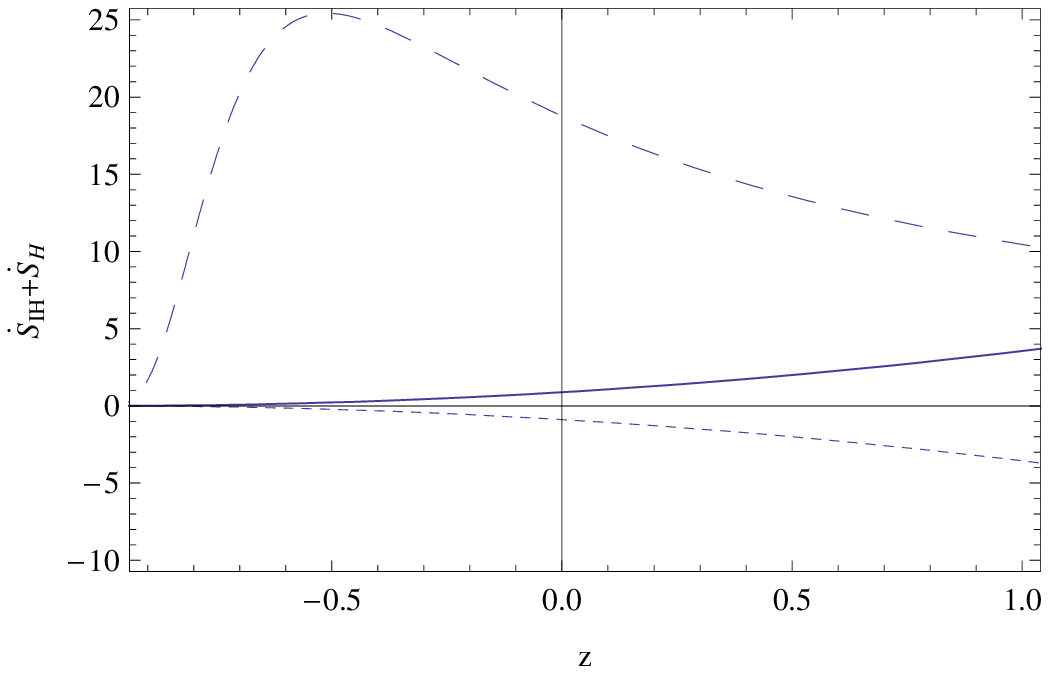}\\
\vspace{1mm} ~~~~~Fig.1~~~~~\\
\vspace{6mm}
\includegraphics[height=2in]{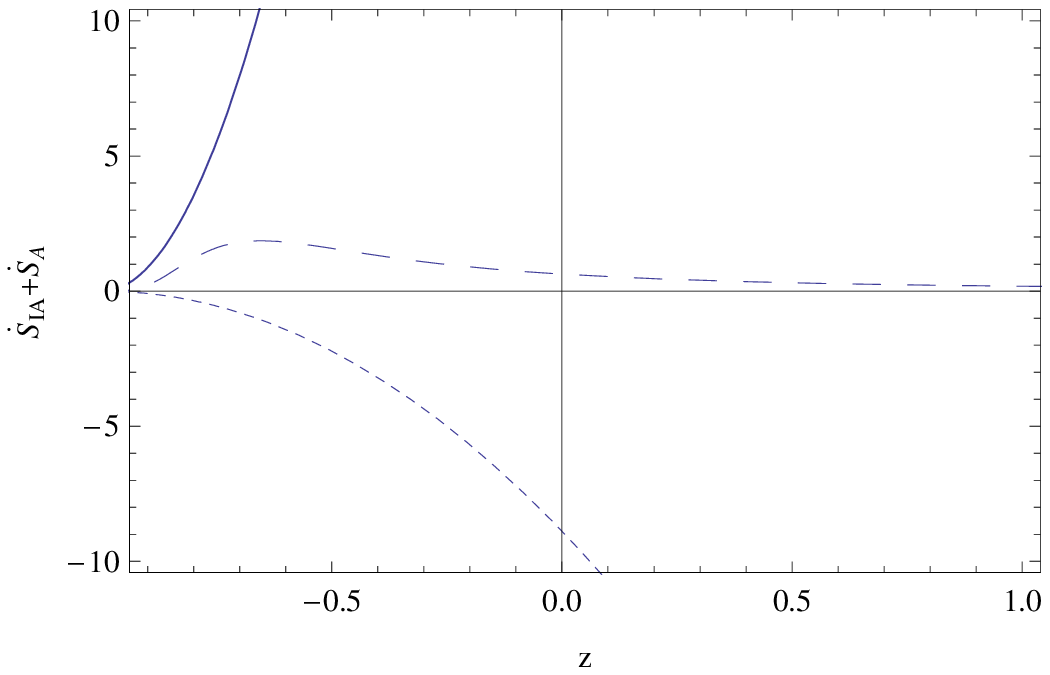}\\
\vspace{1mm} ~~~~~Fig.2~~~~\\

\vspace{6mm}

\includegraphics[height=2in]{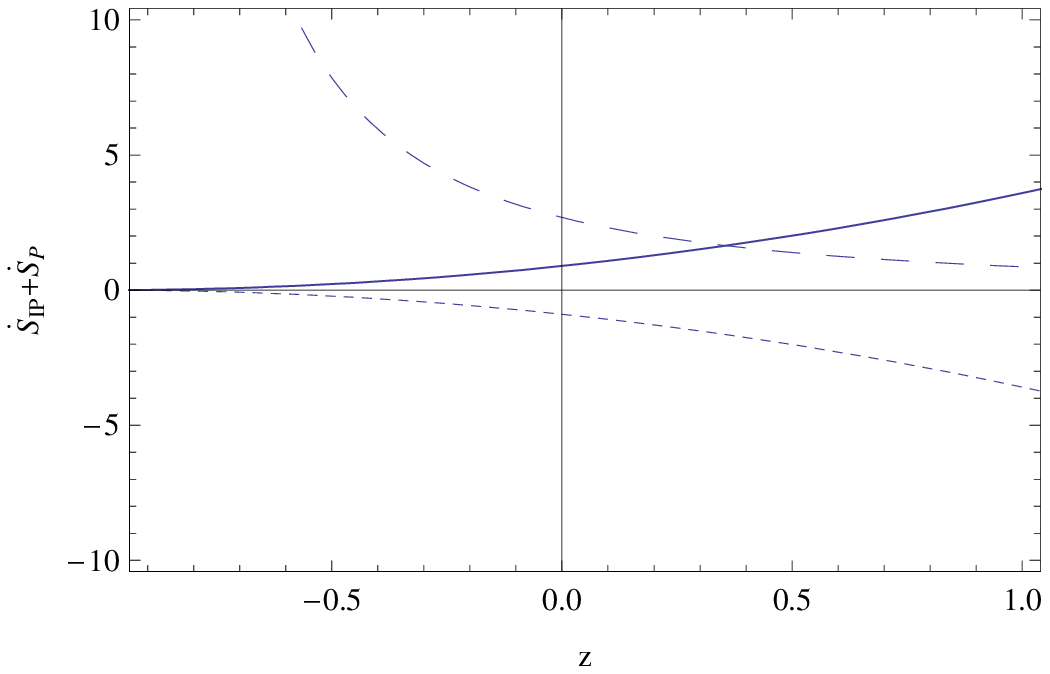}\\
\vspace{1mm} ~~~~~Fig.3~~~~~\\
\vspace{6mm}
\includegraphics[height=2in]{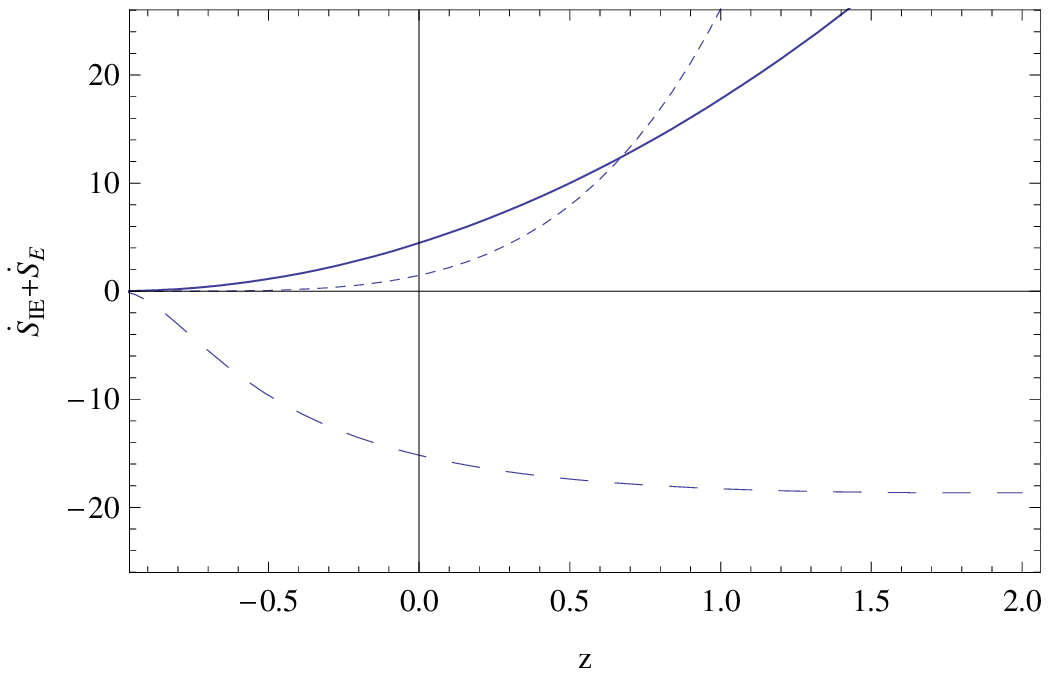}\\
\vspace{1mm} ~~~~~Fig.4~~~~~\\

\vspace{6mm} Figs. 1, 2, 3 and 4 represent respectively the
variations of $(\dot{S}_{IH}+\dot{S}_{H})$,
$(\dot{S}_{IA}+\dot{S}_{A})$, $(\dot{S}_{IP}+\dot{S}_{P})$ and
$(\dot{S}_{IE}+\dot{S}_{E})$ against redshift $z$ for $w=-1/3$ and
$k=0,\pm 1$. The dashed line, dotted line  and filled line
represent for $k=0,~-1$ and $+1$ respectively.

\vspace{6mm}

\end{figure}

Using equations (22) and (28), the total entropy variation on
particle horizon is obtained as

\begin{eqnarray*}
\dot{S}_{IP}+\dot{S}_{P}= - \frac{2 \pi R_{P}^{3}}{G}
\left(\dot{H} - k(1 + z)^{2} + \frac{k^{2}(1 + z)^{4}}{\Lambda
}\right)
\end{eqnarray*}
\begin{equation}
~~~~~~~~~~~~~~~~~~~~~~+ \frac{2 \pi}{G}\left(R_{P} +
\frac{k}{\Lambda R_{P}}\right)(HR_{P} + 1)
\end{equation}

Using equations (23) and (28), the total entropy variation on
event horizon is obtained as

\begin{eqnarray*}
\dot{S}_{IE}+\dot{S}_{E}=\frac{2 \pi R_{E}^{3}}{G}\left(\dot{H} -
k(1 + z)^{2} + \frac{k^{2}(1 + z)^{4}}{\Lambda }\right)
\end{eqnarray*}
\begin{equation}
~~~~~~~~~~~~~~~~~+ \frac{2 \pi}{G}\left(R_{E} + \frac{k}{\Lambda
R_{E}}\right)(HR_{E} - 1)
\end{equation}

The time variations to total entropies on the Hubble, apparent,
particle and event horizons have been drawn against $z$ in figures
1 - 4 respectively for $k=0,\pm 1$. From graphical representations
we make the following conclusions: \\

(a) In figure 1, we see that $(\dot{S}_{IH}+\dot{S}_{H})$ is (i)
always positive for $k=+1$, (ii) always negative for $k=-1$ (iii)
is positive upto certain stage and may be negative at late stage
for $k=0$. So on Hubble horizon, the GSL is satisfied always for
closed universe and for open universe, GSL breaks down. Also for
flat universe, GSL may be satisfied on Hubble horizon but at late
stage ($z<-0.8$) GSL breaks down. \\

(b) In figure 2, we see that $(\dot{S}_{IA}+\dot{S}_{A})$ is (i)
always positive for $k=0$ and $+1$, (ii) is negative for $k=0$. So
on apparent horizon, the GSL is always satisfied for closed and
flat universe. Also for open universe, GSL breaks down. The result
coincides with the work of Jamil et al [20].\\

(c) In figure 3, we see that $(\dot{S}_{IP}+\dot{S}_{P})$ is (i)
always positive for $k=0$ and $+1$ and (ii) always negative for
$k=-1$. So on particle horizon, the GSL is satisfied always for
closed and flat universe and for open universe, GSL breaks down. \\

(d) In figure 4, we see that $(\dot{S}_{IE}+\dot{S}_{E})$ is (i)
always positive for $k=-1$ and $+1$ and (ii) always negative for
$k=0$. So on event horizon, the GSL is satisfied always for
closed and open universe and for flat universe, GSL breaks down. \\

The above conclusions are valid in HL gravity theory with
barotropic fluid solutions ($w=-1/3$). For other types of
solutions, we may obtain similar types of results. So, in HL
gravity, the GSL may be satisfied on different horizons.\\

\section{\normalsize\bf{Generalized second law of Thermodynamics : an effective approach}}

In the effective approach to HL gravity theory, we assume that the
universe contains the matter fluid and dark energy fluid. So the
problem is equivalent to the Einstein's gravity with two fluids.
It can be introduced an effective dark energy defining the energy
density $\rho_{D}$ and pressure $p_{D}$ in the Friedmann equations
$(7)$ and $(8)$ as

\begin{equation}
\rho_{D}\equiv \frac{3 \kappa^{2} \mu^{2} k^{2}}{8(3\lambda
-1)a^{4}} + \frac{3 \kappa^{2} \mu^{2} \Lambda^{2}}{8(3\lambda
-1)}
\end{equation}

and

\begin{equation}
p_{D}\equiv \frac{ \kappa^{2} \mu^{2} k^{2}}{8(3\lambda -1)a^{4}}
- \frac{3 \kappa^{2} \mu^{2} \Lambda^{2}}{8(3\lambda -1)}
\end{equation}

which after the identification $(9)$ and $(10)$, can be written
as,

\begin{equation}
\rho_{D}\equiv \frac{1}{16 \pi G_{c}}\left( \frac{3 k^{2}}{
\Lambda a^{4}} + 3 \Lambda\right)
\end{equation}
and
\begin{equation}
p_{D}\equiv \frac{1}{16 \pi G_{c}}\left( \frac{ k^{2}}{ \Lambda
a^{4}} - 3 \Lambda\right)
\end{equation}

If we assume the matter fluid is conserved then equation (14)
holds and in this case the dark energy conservation equation will
be

\begin{equation}
\dot{\rho}_{D} + 3 H (\rho_{D} + p_{D})=0
\end{equation}

Therefore, the Friedmann equations (12) and (13) turn into the
forms

\begin{equation}
H^{2} + \frac{k}{a^{2}}=\frac{8\pi G_{c}}{3}( \rho + \rho_{D})
\end{equation}

and

\begin{equation}
\dot{H} + \frac{3}{2}H^{2} + \frac{k}{2a^{2}}= -4 \pi G_{c}(p +
p_{D})
\end{equation}

As in the previous section, we find the variations of entropies
for matter fluid and dark energy fluid respectively as

\begin{equation}
\dot{S}=\frac{4\pi}{T_{X}}(\rho + p) R_{X}^{2}( \dot{R}_{X} - H
R_{X})
\end{equation}

and

\begin{equation}
\dot{S}_{D}=\frac{4\pi}{T_{X}}( \rho_{D} + p_{D})R_{X}^{2}(
\dot{R}_{X} - H R_{X})
\end{equation}

Since the effective gravitational sector is now the standard
general relativity, the horizon entropy will be given by

\begin{equation}
S_{X}=\frac{\pi R_{X}^{2}}{G}
\end{equation}

where IR limit $(\lambda=1)$ which allows us to simplify
$G_{c}=G$. Therefore, taking derivative, we get

\begin{equation}
\dot{S}_{X}=\frac{2 \pi R_{X} \dot{R}_{X}}{G}
\end{equation}

Adding the relations $(47)$, $(48)$ and $(50)$, the rate of change
of total entropy is obtained as

\begin{eqnarray*}
\dot{S}_{tX}= \dot{S}_{D} + \dot{S} + \dot{S}_{X}=8 \pi^{2}
R_{X}^{3}( \dot{R}_{X} - H R_{X})(\rho + p + \rho_{D} + p_{D})
\end{eqnarray*}
\begin{equation}
~~~~~~~~~~~~~~~~~~~~~~~~~~+ \frac{2 \pi R_{X} \dot{R}_{X}}{G}
\end{equation}

From the field equations $(45)$ and $(56)$ we get

\begin{equation}
\rho + p + \rho_{D} + p_{D}=-\frac{1}{4 \pi G_{c}}\left(\dot{H} -
\frac{k}{a^{2}}\right)
\end{equation}

Substituting  expressions for different horizons  with radius
$R_{X}$ in $(51)$
we get the following results:\\

For Hubble horizon,

\begin{equation}
\dot{S}_{tH}= \frac{2 \pi}{GH^{3}}\left(\frac{\dot{H}}{H^{2}} +
1\right)\left(\dot{H} - \frac{k}{a^{2}}\right) - \frac{2 \pi
\dot{H} }{G H^{3}}
\end{equation}

For apparent Horizon,

\begin{equation}
\dot{S}_{tA}=\frac{2\pi}{G}R_{A}^{6}H\left(\dot{H} -
\frac{k}{a^{2}}\right)^{2}
\end{equation}

For particle horizon,

\begin{equation}
\dot{S}_{tP}= - \frac{2 \pi R_{P}^{3}}{G}\left(\dot{H} -
\frac{k}{a^{2}}\right) + \frac{2 \pi R_{P} }{G}(H R_{P} + 1 )
\end{equation}

For event horizon,

\begin{equation}
\dot{S}_{tE}= \frac{2 \pi R_{E}^{3}}{G}\left(\dot{H} -
\frac{k}{a^{2}}\right) + \frac{2 \pi R_{E} }{G}(H R_{E} -1 )
\end{equation}

\begin{figure}
\includegraphics[height=2in]{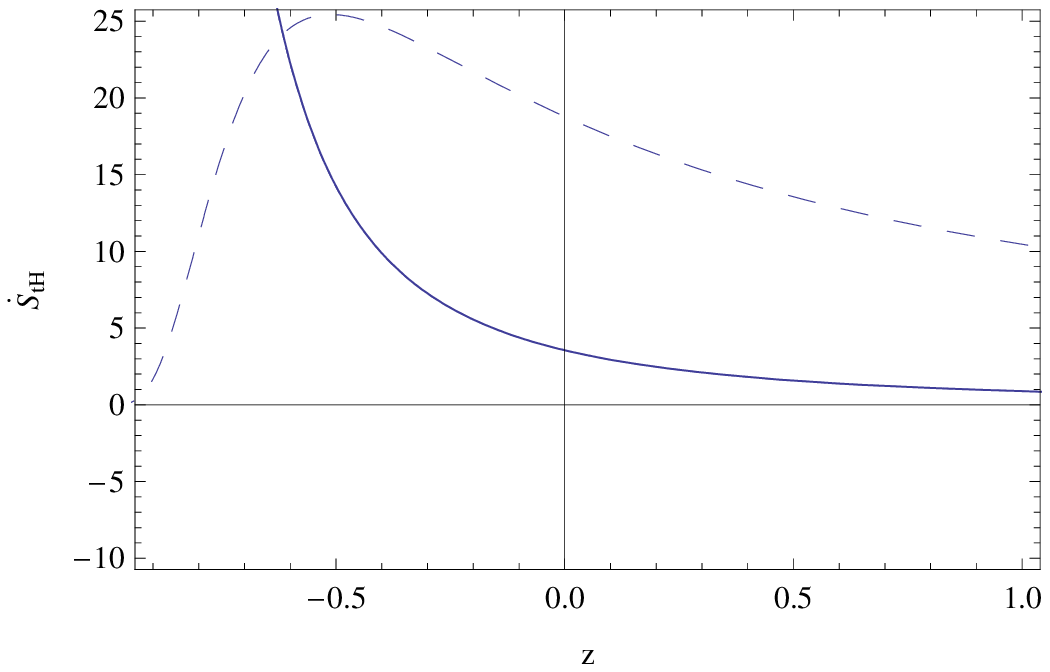}\\
\vspace{1mm} ~~~~~Fig.5~~~~~\\
\vspace{6mm}
\includegraphics[height=2in]{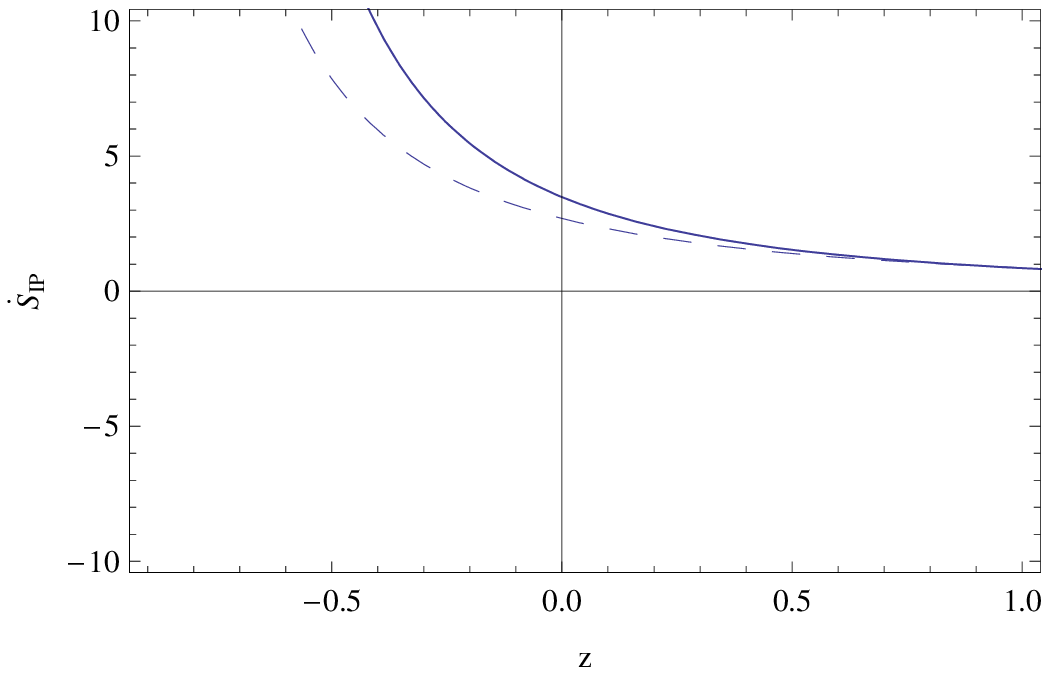}\\
\vspace{1mm} ~~~~~Fig.6~~~~~\\

\vspace{6mm}

\includegraphics[height=2in]{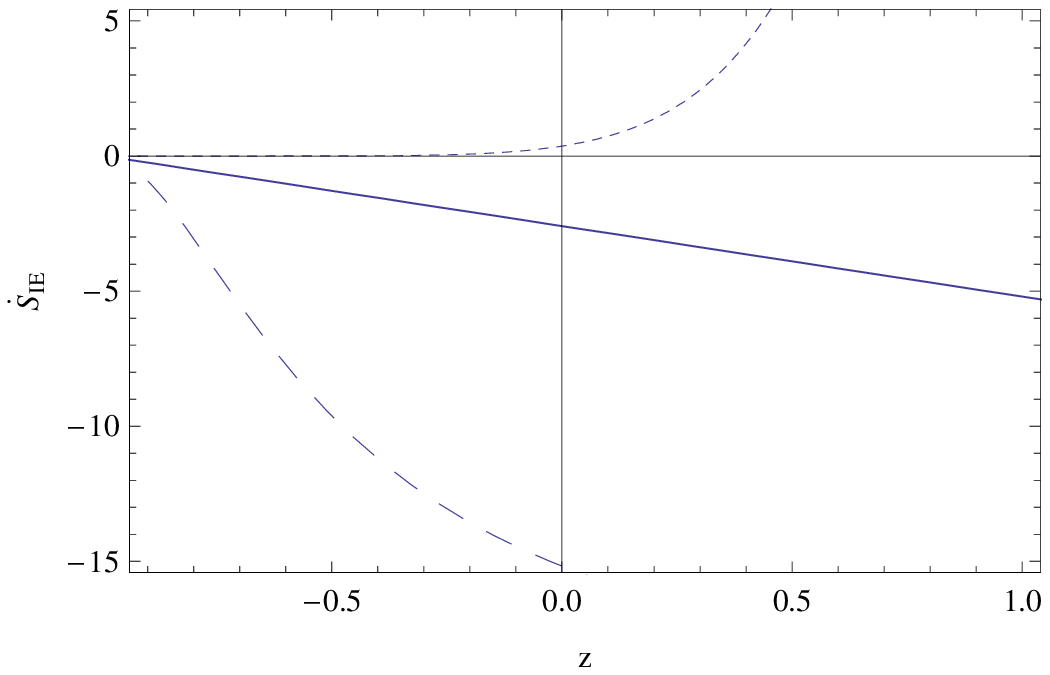}\\
\vspace{1mm} ~~~~~Fig.7~~~~~~\\

\vspace{6mm} Figs. 5, 6 and 7 represent respectively the
variations of $\dot{S}_{tH}$, $\dot{S}_{tP}$ and $\dot{S}_{tE}$
against redshift $z$ for $w=-1/3$ and $k=0,\pm 1$. The dashed
line, dotted line  and filled line represent for $k=0,~-1$ and
$+1$ respectively.\\

\end{figure}

From (53), we see that $\dot{S}_{tA}\ge 0$ always for $k=0,\pm 1$.
So GSL is satisfied on apparent horizon for flat, open and closed
FRW universe in Ho$\check{\text r}$ava-Lifshitz gravity. But
$\dot{S}_{tH},~\dot{S}_{tP}$ and $\dot{S}_{tE}$ will be positive
if the r.h.s. of equations (53), (55) and (56) are positive. So
validity of GSL on Hubble, particle and event horizons depend on
the values of $H,~\dot{H}$ and horizons radii for flat, open and
closed FRW universe in Ho$\check{\text r}$ava-Lifshitz gravity.
The expressions of $H,~\dot{H}, R_{H}, R_{P}$ and $R_{E}$ for
barotropic fluid model with HL dark energy are given in equations
(16) and (32) - (35). The time variations of total entropies on
Hubble, particle and event horizons are presented in figures 5 - 7
for $k=0,\pm 1$. From graphical representations we make the following
conclusions: \\

(a) In figure 5, we see that $\dot{S}_{tH}$ is (i) always positive
for $k=-1$ and $+1$, (ii) is positive upto certain stage and may
be negative at late stage for $k=0$. So on Hubble horizon, the GSL
is satisfied always for open and closed universe. Also for flat
universe, GSL may be satisfied on Hubble horizon but
at late stage ($z<-0.8$) GSL breaks down. \\

(b) In figure 6, we see that $\dot{S}_{tP}$ is always positive for
$k=0,\pm 1$ and $+1$. So on particle horizon, the GSL is satisfied
always for open, closed and flat universe. \\

(c) In figure 7, we see that $\dot{S}_{tE}$ is always negative for
$k=0,\pm 1$. So on event horizon, the GSL cannot be satisfied for
open, closed and flat universe. \\

The above conclusions are valid in HL gravity theory with
barotropic fluid solutions ($w=-1/3$). For other types of
solutions, we may obtain similar types of results. So, in HL
gravity, the GSL may or may not be satisfied on different
horizons.\\

\section{\normalsize\bf{Discussions}}

In this work, we have investigated the validity of GSL of
thermodynamics in a universe (open, closed and flat) governed by
Ho$\check{\text r}$ava-Lifshitz gravity. If the universe contains
barotropic fluid the corresponding solutions have been obtained.
Consider the universe as a thermodynamical system bounded by
horizons. The validity of GSL have been examined by two
approaches: (i) robust approach and (ii) effective approach. In
robust approach, we have considered the universe contains only
matter fluid and the effect of the gravitational sector of HL
gravity was incorporated through the modified black hole entropy
on the horizon. The general prescription for validity of GSL have
been discussed. But we cannot draw any definite conclusion for
validity of GSL in open, closed and flat models. So graphical
experiments have been investigated for final conclusion. The total
variations of entropies on Hubble, apparent, particle and event
horizons have been presented in figures 1 - 4. These figures show
that on Hubble horizon, the GSL is satisfied always for closed
universe and for open universe, GSL breaks down. Also for flat
universe, GSL may be satisfied on Hubble horizon but at late stage
($z<-0.8$) GSL breaks down. On apparent horizon, the GSL is always
satisfied for closed and flat universe. Also for open universe,
GSL breaks down. On particle horizon, the GSL is satisfied always
for closed and flat universe and for open universe, GSL breaks
down. On event horizon, the GSL is satisfied always for closed and
open universe and for flat universe, GSL breaks down.\\

Effective approach is that all extra information of HL gravity
into an effective dark energy fluid and so we consider the
universe contains matter fluid plus this effective fluid. This
approach is essentially same as the Einstein's gravity theory. In
this situation, we have obtained the general conditions for
validity of GSL in open, closed and flat model of the universe. It
has been seen that on the apparent horizon, the GSL is always
valid. But for other horizons, it may or may not be valid. The
total variations of entropies on Hubble, particle and event
horizons have been presented in figures 5 - 7. From figures, we
see that on Hubble horizon, the GSL is satisfied always for open
and closed universe. Also for flat universe, GSL may be satisfied
on Hubble horizon but at late stage ($z<-0.8$) GSL breaks down. On
particle horizon, the GSL is satisfied always for open, closed and
flat universe. On event horizon, the GSL cannot be satisfied for
open, closed and flat universe. \\

{\bf Acknowledgement:}\\

The authors are thankful to IUCAA, Pune, India for warm
hospitality where part of the work was carried out.\\

{\bf References:}\\
\\
$[1]$ P. Ho$\check{\text r}$ava, arXiv:0811.2217.\\
$[2]$ P. Ho$\check{\text r}$ava, {\it JHEP} {\bf 03} 020 (2009).\\
$[3]$ P. Ho$\check{\text r}$ava, {\it Phys. Rev. D} {\bf 79}
084008 (2009).\\
$[4]$ P. Ho$\check{\text r}$ava, {\it Phys. Rev. Lett.} {\bf 102}
161301 (2009).\\
$[5]$ R. -G. Cai, B. Hu and H. B. Zhang, {\it Phys. Rev. D} {\bf
80} 041501 (2009).\\
$[6]$ G. Calcagni, {\it JHEP} {\bf 09} 112 (2009).\\
$[7]$ E. Kiritsis and G. Kofinas, {\it Nucl. Phys. B} {\bf 821}
467 (2009).\\
$[8]$ T. Takahashi and J. Soda, {\it Phys. Rev. Lett.} {\bf 102} 231301 (2009).\\
$[9]$ S. Mukohyama, {\it JCAP} {\bf 0906} 001 (2009).\\
$[10]$ H. Lu, J. Mei and C. N. Pope, {\it Phys. Rev. Lett.} {\bf 103} 091301 (2009).\\
$[11]$ R. G. Cai, L. M. Cao and N. Ohta, {\it Phys. Rev. D} {\bf
80} 024003 (2009); R. G. Cai, Y. Liu and Y. W. Sun, {\it JHEP} {\bf 0906} 010 (2009).\\
$[12]$ G. Leon and E. N. Saridakis, {\it JCAP}
{\bf 0911} 006 (2009); M. Minamitsuji, {\it Phys. Lett. B} {\bf 684} 194 (2010).\\
$[13]$ S. Mukhohyama, K. Nakayama, F. Takahashi and S. Yokoyama,
{\it Phys. Lett. B} {\bf 679} 6 (2009); T. Takahashi and J. Soda,
{\it Phys. Rev. Lett.} {\bf 102} 231301 (2009).\\
$[14]$ M. I. Park, {\it JCAP} {\bf 1001} 001 (2010);
M. R. Setare and M. Jamil, {\it JCAP} {\bf 1002} 010 (2010).\\
$[15]$ S. S. Kim, T. Kim and Y. Kim, {\it Phys. Rev. D} {\bf 80}
124002 (2009); K. Izumi and S. Mukohyama, {\it Phys. Rev. D} {\bf 81} 044008 (2010).\\
$[16]$ S. Dutta and E. N. Saridakis, {\it JCAP} {\bf 1005} 013
(2010).\\
$[17]$ C. Charmousis, G. Niz, A. Padilla and P. M. Saffin, {\it
JHEP} {\bf 0908} 070 (2009); T. P. Sotiriou, M. Visser and S.
Weinfurtner, arXiv:0905.2798 [hep-th]; C. Bogdanos and E. N.
Saridakis, {\it Class. Quant. Grav.} {\bf 27} 075005 (2010).\\
$[18]$ R. B. Mann, {\it JHEP} {\bf 0906} 075 (2009); G. Bertoldi,
B. A. Burrington and A. Peet, {\it Phys. Rev. D} {\bf 80} 126003
(2009); R. G. Cai, L. M. Cao and N. Ohta, {\it Phys. Lett. B} {\bf 679} 504 (2009).\\
$[19]$ R. -G. Cai and N. Ohta, {\it Phys. Rev. D} {\bf 81} 084061
(2010); N. Mazumder and S. Chakraborty, arXiv: 1003.1606[gr-qc];
M. Jamil, A. Sheykhi and M. U. Farooq, arXiv:1003.2093[hep-th]; A.
Wang and Y. Wu, {\it JCAP} {\bf 0907} 012 (2009); Q. -J. Cao, Y.
-X. Chen and K. -N. Shao, arXiv:1001:2597[hep-th].\\
$[20]$ M. Jamil, E. N. Saridakis and M. R. Setare, arXiv:
1003.0876 [hep-th].\\
$[21]$ R. L. Arnowitt, S. Deser and C. W. Misner, {\it The
Dynamics of General Relativity} appeared as Chapter 7, pp.
227-264, in {\it gravitation: an introduction to current
research}, L. Witten, ed. (Wiley, New York, 1962), arXiv:
gr-qc/0405109.\\
$[22]$ S. M. Carroll and E. A. Lim, {\it Phys. Rev. D} {\bf 70}
123525 (2004).\\
$[23]$ R. G. Cai, L. M. Cao and N. Ohta, {\it Phys. Lett. B} {\bf
679} 504 (2009);  R. G. Cai and N. Ohta, {\it Phys. Rev. D} {\bf 81} 084061 (2010).\\

\end{document}